\pdfoutput=1

\RequirePackage{fix-cm}

\documentclass[smallextended]{svjour3}       

\usepackage{makeidx}
\usepackage{graphicx}
\usepackage{url}
\usepackage{algpseudocode}
\usepackage[ruled,section]{algorithm}
\usepackage{booktabs}
\usepackage{enumitem}
\usepackage{multirow}
\usepackage{amsmath}
\usepackage{amssymb}
\usepackage{booktabs}
\usepackage{pbox}
\usepackage{xcolor}
\usepackage{natbib}
\usepackage{color, colortbl}
\definecolor{LGray}{rgb}{.8,.8,.8}

\smartqed  

\begin{document}

\title{General factorization framework for context-aware recommendations}


\author{Bal\'azs Hidasi \and Domonkos Tikk}


\institute{Bal\'azs Hidasi \and Domonkos Tikk \at
              Gravity R\&D, \\
              \email{\{balazs.hidasi,domonkos.tikk\}@gravityrd.com}
           \and
           Bal\'azs Hidasi \at
              Budapest University of Technology and Economics
}


\maketitle

\newcommand{\dom}{\mathrm{dom}}

\begin{abstract}
Context-aware recommendation algorithms focus on refining recommendations by considering additional information, available to the system. This topic has gained a lot of attention recently. Among others, several factorization methods were proposed to solve the problem, although most of them assume explicit feedback which strongly limits their real-world applicability. While these algorithms apply various loss functions and optimization strategies, the preference modeling under context is less explored due to the lack of tools allowing for easy experimentation with various models. As context dimensions are introduced beyond users and items, the space of possible preference models and the importance of proper modeling largely increases.

In this paper we propose a General Factorization Framework (GFF), a single flexible algorithm that takes the preference model as an input and computes latent feature matrices for the input dimensions. GFF allows us to easily experiment with various linear models on any context-aware recommendation task, be it explicit or implicit feedback based. The scaling properties makes it usable under real life circumstances as well.

We demonstrate the framework's potential by exploring various preference models on a 4-dimensional context-aware problem with contexts that are available for almost any real life datasets. We show in our experiments -- performed on five real life, implicit feedback datasets -- that proper preference modelling significantly increases recommendation accuracy, and previously unused models outperform the traditional ones. Novel models in GFF also outperform state-of-the-art factorization algorithms.

We also extend the method to be fully compliant to the Multidimensional Dataspace Model, one of the most extensive data models of context-enriched data. Extended GFF allows the seamless incorporation of information into the factorization framework beyond context, like item metadata, social networks, session information, etc. Preliminary experiments show great potential of this capability.

\end{abstract}

\section{Introduction}\label{sec:intro}

In this paper we propose a (1) general factorization framework (GFF) that (2) works also on implicit feedback data; (3) integrates recommendation context into the model; (4) and is flexible enough to employ the various underlying relationships of dimensions allowing us to adequately model preferences. We first argue why we consider these design points important.

Recommender systems \citep{RicciRSH} are information filtering tools that help users in information overload to find relevant content. Here we focus on the class of latent factor based collaborative filtering (CF) methods that gained popularity due to their good \emph{accuracy} and \emph{scalability} \citep{KorenRSH11}. They capture the users' preferences by uncovering latent features that explain the observed user--item ratings using factor models.

In most practical scenarios, however, users do not rate content/items explicitly: one can only observe the users' interactions\footnote{User purchased an item or viewed an product page, etc. Interactions also called events or transactions.} with items---retrieved from web logs, for instance---as they use the system. This type of feedback is termed implicit feedback, also called one-class CF in the literature, and contains unary data, i.e. recorder user--item interactions. 

Implicit feedback data contains less information on user preferences than explicit feedback.
Explicit feedback requires the active contribution of the users to state their preferences on items they consumed or familiar with; thus it can directly encode both the positive and negative opinions. Implicit feedback is less accurate and negative feedback is missing. Firstly, user interactions can only be \emph{interpreted} as positive feedback; this can be inaccurate when, e.g. user is disappointed with a purchased item, or clicks on an article due to clickbaiting. Secondly, direct negative preferences are completely missing: the lack of an interaction typically means that the user was not aware of the item's existence.\footnote{We use the classic notion of implicit and explicit feedback here. In some cases explicit feedback can also be positive only, called also unary rating, typical is the voting scenario, such as Facebook likes, or Google's $+1$, \cite{RicciRSH11}. However our focus is the easily collectable implicit feedback, that is unary data. While one can (and must) infer negative signs of preference from such data, e.g., by considering missing feedback or using additional information such as time spent on page, negative and positive preferences are not explicitly distinguished.} The fact, however, that implicit feedback is always available, highlights the importance of methods working on such data for real-world recommendation applications.

Context-aware recommendation systems (CARS) refine recommendations by considering additional information, available to the system. They extend the dualistic user--item modeling concept and consider additional information that may influence the user preferences at recommendation. Such data are together termed \emph{contextual information}, or briefly \emph{context} \citep{AdomaviciusRecsys08}. One class of CARS uses latent factor methods (see e.g. \citet{KaratzogluRecsys10,RendleWSDM2010,itals_ecml,Shi_tfmap,rendle2012factorization}). Most of the latent factor based CARS however work only on explicit feedback problems that strongly limits their real-world applicability.



Factorization methods are characterized by three components, see e.g. \citep{BellkorICDM07,brismf,Salak08}. (1) A loss function that is to be minimized by the algorithm. The loss is a function of predicted preferences (or ratings) and usually (but not necessarily) contains the difference between actual and predicted preferences (or ratings). (2) An optimization method that iteratively optimizes the value of the latent factors in order to minimize the loss function. (3) A preference model that describes how preferences are estimated. For example: BRISMF \citep{brismf} optimizes for minimal root mean squared error (RMSE) loss, with a stochastic gradient descent (SGD) optimizer under a preference model where the preference of a user on an item is predicted by the dot product of the user's and the item's feature vector; BPR \citep{bpr} optimizes for maximal BPR criteria using SGD optimization strategy, with the same preference model as BRISMF; iTALS \citep{itals_ecml} optimizes for a weighted RMSE loss by using ALS optimization strategy with a preference model where the preference of a user on an item under context is the N-way dot product of the user's the item's and the context-states' feature vectors\footnote{More precisely: the sum of elements in the elementwise product of corresponding vectors.}; Factorization Machines \citep{rendle2012factorization} optimizes for minimal RMSE loss, with one of three optimization strategies (SGD, coordinate descent or Bayesian inference using Markov Chain Monte Carlo (MCMC)) with a pairwise interaction preference model, i.e. the sum of dot products between the feature vectors of every pairs of entities (e.g. user with item, user with context-state and item with context-state).

Different CARS apply various loss functions and optimization strategies. However preference modeling under context is less explored. Most methods use either the N-way or the pairwise interaction model. In the former, the preference is predicted by an N-way dot product between interacting entities; the latter calculates the model as the sum of pairwise dot products between every pair of interacting entities. As additional (context) dimensions are introduced beyond users and items, the space of possible preference models and the importance of proper modeling largely increase. We argue that the lack of proper exploration of this area is due to the lack of flexible tools in which one can experiment with various models without being required to implement a specific algorithm for each model. We therefore created the General Factorization Framework (GFF), a single, flexible algorithm that takes the preference model as an input and computes latent feature matrices for the input dimensions. GFF allows us to easily experiment with various linear models on any context-aware recommendation task, be it explicit or implicit feedback based. We believe that GFF opens up a new research path in preference modeling under context.

The following properties were important at the design of GFF.
\begin{enumerate}
\item No restriction on context: GFF works on any context-aware recommendation problem independently of the number and the meaning of context dimensions.
\item Large preference model class: the only restriction on the preference model is that it must be linear in the dimensions of the problem\footnote{Meaning that a dimension can not directly interact with itself in the model}
. This intuitive restriction does not restrict the applicability to real-world problems.
\item Data type independence: besides the practically more useful implicit case, explicit problems can be also addressed by simply changing the weighting scheme in the loss function.
\item Flexibility: the weighting scheme of GFF is very flexible, enabling to incorporate extra knowledge through the weights such time decay, dwell time dependent weighting, missing not at random hypotheses and more.
\item Scalability: GFF scales well both in terms of the number of interactions in the training set and in the number of features. This makes it applicable in real life recommender systems.
\end{enumerate}

The rest of the paper is organized as follows. Section~\ref{sec:datamodel} introduces dataspace models for context-enriched data. Building on this, the basic version of GFF is introduced in Section~\ref{sec:gff-basic}. In Section~\ref{sec:models} we demonstrate the usefulness of GFF by experimenting with different models on a 4 dimensional context-aware problem. 
The results clearly imply that proper preference modeling is important is this field.
We compare the results of some models in GFF to state-of-the-art methods in Section~\ref{sec:comp-other}. GFF is further extended in Section~\ref{sec:gff-extended} to be fully compliant to the Multidimensional Dataspace Model. Extended GFF allows the seamless incorporation of information into the factorization framework beyond context, like item metadata, social networks, session information, etc. Preliminary experiments show great potential of this capability. Finally, Section~\ref{sec:conclusion} summarizes this work and hints on future research. 

\section{Data model}\label{sec:datamodel}
In this section we briefly review data models for the representation of context-aware data. The focus is on the representation of the input, that is users, items, context; the target attribute (e.g. rating, preference) can be added in a straightforward way. One of the most extensive data models for this task is the Multidimensional Dataspace Model (MDM,  \citet{AdomaviciusACMTIS05}). In MDM the dataspace is the Cartesian product of $N_D$ dimensions: $DS=D_1\times D_2\times\cdots\times D_{N_D}$. Each dimension contains one or more attributes: $D_i=A_{i,1}\times A_{i,2}\times\cdots\times A_{i,N_i}$. The data model is very similar to that of relational databases. It is usually also required that the values of an attribute come from a set of atomic and nominal attributes. Therefore continuous variables should be discretized and the order between attribute values is disregarded. The data -- usually in the form of transactions -- is the subset of every possible combination of the attribute values of all attributes of all dimensions.

We give an example for representing data in MDM. Let $D_1=U$ be the dimension for users, $D_2=I$ the dimension for items, and $D_3=L$ the dimension for locations, thus the dataspace is every possible combination of users, items and locations, i.e. $DS=U\times I\times L$. Let us describe the users by their ID, gender and age; the items by their ID and genres; and the location by the city. Note the following: (1) The data model does not require using the IDs for users/items. However in the classical recommendation scenario the system recommends individual items to individual users. Therefore IDs should be present to distinguish them. If the subject of the recommendation is not an item but one item property, the ID can be omitted. (2) If an item can belong to only one genre, then the item dimension has one attribute that contains this information. If an item has multiple genres then either the combination of genres are the attribute values for a single genre attribute or a binary attributes are required for each genre (e.g. IsAction, IsComedy, etc.) that contain one if the item belongs to that genre.

Factorization methods usually use a simplified version of MDM. There are several ways to simplify MDM, here we review the major ones.

Generic factorization methods -- such as Factorization Machines (FM) \citep{rendle2012factorization} or iTALS(x) (\citet{itals_ecml,italsx_infocomm}) -- restrict the number of attributes to one per dimension, however, do not limit the number of dimensions. We refer to this data model as Single Attribute MDM (SA-MDM). Most information can be equally represented in SA-MDM as MDM by just ignoring the grouping of attributes by the dimensions. The main conceptual difference is that interactions between attributes of the same dimension (e.g. item IDs and item genres) cannot be captured. By ``converting'' all attributes to dimensions we lose the information of this grouping and thus assume extra interactions. This may result in much more interactions (and therefore complexity), especially if multi-valued attributes, like genre or category, is decomposed to many binary attributes.

The data model can be further simplified by setting a limit on the number of dimensions as well. Matrix factorization (e.g. \citet{HuICDM08,bpr,brismf}) limits the number of dimensions to two (one for users, one for items) and several tensor factorization methods work on only three dimensional data (e.g. \citet{RendleWSDM2010,Shi_tfmap}).

An other interesting variant of MDM is when the number of dimensions is fixed, but the number of attributes in a dimension is not. Prominent examples using such data model are SVDFeature \citep{Chen12JMLR}, SVD++ \citep{Koren2008KDD} and NSVD1 \citep{paterek2007improving}. They use two fixed dimensions: users and items. SVDFeature sets no restrictions on the number and meaning of attributes for neither the users nor the items. SVD++ requires one of the dimensions to contain a single ID attribute only while the other dimension consists of an ID and several other attributes. Usually the user dimension is restricted to the ID and the additional attributes in this case are binary attributes for all item IDs that are set to one if the user intereacted with the given item. NSVD1 also restricts one of the dimensions to an ID attribute, while the other consists of binary entities of descriptor entities. The descriptor entities are either metadata tokens or users that rated the given item.

GFF is designed to be fully compliant with MDM. The framework has two levels: basic GFF builds on SA-MDM (see Section~\ref{sec:gff-basic}), while the extended version also incorporates multiple attributes per dimensions (see Section~\ref{sec:gff-extended}.

\section{Basic GFF}\label{sec:gff-basic}
GFF is a general modeling framework --- inspired by the latent factor CF approach --- which (1) efficiently integrates context data into the preference model; (2) allows experimentation with non-traditional models for more accurate preference estimation.

The basic framework relies on SA-MDM (see Section~\ref{sec:datamodel}). In recommendation problems, the main goal is the modeling of user preferences on items, therefore one dimension is dedicated for the \emph{users} and one dedicated for the \emph{items}. We use one ID attribute in these dimensions. Other dimensions contain context data that helps modeling user preferences. Context can be the location or time of the interaction, the device on which the interaction was performed, or any other parameters that may influence the user preference, including weather, referral's link, search keyword, etc. Since SA-MDM is used, each context dimension contains exactly one attribute. The preference model is solely learnt from sample \emph{events} (also called transactions).

Inspired by factorization methods, we assign a feature vector of length $K$ to each possible value of each attribute. We refer to these values as entities. For instance, the possible user IDs are entities. Therefore each attribute is represented as a feature matrix ($M^{(i)}\in \mathbb{R}^{K\times S_i}$, where $S_i$ is the number of entities in the $i^{\rm{th}}$ dimension), assembled from the feature vectors of entities of the attribute. Since each dimension consists of exactly one attribute, dimensions are also represented by this feature matrix.

SA-MDM compliant data can be arranged into an $N_D$ dimensional tensor $R$. The values in the tensor are the preferences for the given combination of entities (i.e. a user-item-context combination). In case of explicit feedback data, the preferences are ratings. Typically the data space is very sparse, few ratings are observed, others are missing. Our focus is the implicit case and therefore $R$ is filled with binary preference information: if a combination of entities occurred in the training data then the corresponding cell is set to 1, otherwise to 0.
\begin{equation}
    r_{i_1,\ldots,i_{N_D}}=
\begin{cases}
    1,& \text{if } t_{i_1,\ldots,i_{N_D}}\in T\\
    0,              & \text{otherwise}
\end{cases}
\end{equation}

Since the missing feedback is clearly a weaker signal of negative preference than the presence of positive feedback we construct a $\mathcal{W}(i_1,\ldots,i_{N_D})$ weight function that assigns a real value to every possible entity combination. In practice, the construction of $\mathcal{W}(\cdot)$ depends on the problem, and can also affect the complexity of the training. In order to be able to train the model efficiently we restrict $\mathcal{W}(\cdot)$ as follows:
\begin{equation}\label{eq:weight}
\begin{gathered}
    \mathcal{W}: ({i_1,\ldots,i_{N_D}}) \rightarrow \mathbb{R}\\
\mathcal{W}({i_1,\ldots,i_{N_D}})=\begin{cases}
    w^1(i_1,\ldots,i_{N_D}) \gg w^0({i_1,\ldots,i_{N_D}}), &\text{if } t_{i_1,\ldots,i_{N_D}}\in R\\
    w^0(i_1,\ldots,i_{N_D})=\prod_{j=1}^{N_D}{\left(\mu^{(j)}v^{(j)}_{i_j}+\gamma^{(j)}\right)}, &\text{otherwise}
\end{cases}
\end{gathered}
\end{equation}
Where $w^1(i_1,\ldots,i_{N_D})$ is the weight of entity combinations of the training set and $w^0(i_1,\ldots,i_{N_D})$ is the weight of missing entity combinations. Both weight functions depend on the actual entities. Note that we require $w^0(\cdot)$ to be factorized by the dimensions. $v^{(j)}_{i_j}$ is a weight for the $(i_j)^{\rm{th}}$ entity in the $j^{\rm{th}}$ dimension. This weight can depend on any property of the entity. $\mu^{(j)}$ and $\gamma^{(j)}$ are constants for the $j^{\rm{th}}$ dimension. Therefore the weight by a given dimension can be either a constant or depend on a property of the actual entity. Although this sufficiently generic weight function class enables using different weighting schemes, we leave the exploration of its effect to future research.

For the sake of simplicity, in this paper we use a simple weight function by setting $\mu^{(j)}=0$ and $\gamma^{(j)}=1$ for all $j$, and setting $w^1(i_1,\ldots,i_{N_D})=\alpha\cdot \#(i_1,\ldots,i_{N_D})$. That is $w^0(\cdot)=w_0=1$ for every entity combination and $w^1(\cdot)$ is proportional with the number of occurrences of said combination in the training set. This basic weighting assumes that entity combinations are missing at completely random \citep{MissingData} and that it is more important to accurately predict for entity combinations with actual feedback than for ones with no feedback. This weighting scheme is the generalization of the concept introduced in \citep{HuICDM08}.\footnote{Note that by setting $w^0=0$ and $w^1=1$ and using ratings in $R$ we get the standard explicit setting in $N_D$ dimensions.}
\begin{equation}\label{eq:simpleweight}
\begin{gathered}
\mathcal{W}(i_1,\ldots,i_{N_D})=\begin{cases}
    w^1(i_1,\ldots,i_{N_D})=\alpha\cdot\#(i_1,\ldots,i_{N_D}) \gg w^0_{i_1,\ldots,i_{N_D}}, &\text{if } t_{i_1,\ldots,i_{N_D}}\in R\\
    w^0(i_1,\ldots,i_{N_D})=w_0=1, &\text{otherwise}
\end{cases}
\end{gathered}
\end{equation}

We define the loss as the weighted sum of squared loss:\footnote{We omit regularization for clearer presentation, but $\ell_2$ regularization is used in the actual algorithm.}
\begin{equation}\label{eq:loss}
L=\sum_{i_1=1,\ldots,i_{N_D}=1}^{S_1,\ldots,S_{N_D}}{\mathcal{W}(i_1,\ldots,i_{N_D})(\hat{r}_{i_1,\ldots,i_{N_D}}-r_{i_1,\ldots,i_{N_D}})^2}
\end{equation}

The main novelty in GFF is that the preference model, i.e. the computation of $\hat{r}_{i_1,\ldots,i_{N_D}}$ is an input of the algorithm. This allows us to experiment with \emph{any linear models} beyond the usual ones. In the general framework, a preference model is a linear model of the feature vectors such that: (1) a model consists of sums of Hadamard (or elementwise) products; (2) each product contains at least two feature vectors; (3) in a product each feature vector belongs to a different attribute (linearity); (4) constant importance weights can be applied to each product.\footnote{Omitted from the deduction for clearer presentation.}
\begin{equation}\label{eq:model}
\begin{aligned}
    \hat{r}_{i_1,\ldots,i_{N_D}}=1^T\big(&M^{(\sigma_1)}_{\pi_1}\circ\ldots\circ M^{(\sigma_{p_1})}_{\pi_{p_1}}+\ldots+M^{(\sigma_{p_{q-1}+1})}_{\pi_{p_{q-1}+1}}\circ\ldots\circ M^{(\sigma_{p_q})}_{\pi_{p_q}}\big)
\end{aligned}
\end{equation}
where $\sigma_k\in[1\ldots {N_D}]$ and $\pi_k=i_j$ if $\sigma_k=j$. Biases can be included in the feature vectors and are not presented here separately due to clearer presentation. The model basically consists of selected interactions between members of a subset of dimensions.

\subsection{Training with ALS(-CG)}
Recall that the framework is designed to work also for implicit feedback, thus we need an optimization method that can efficiently handle the implicit setting. Methods that work for the explicit case can not be applied directly for the implicit case due to scalability issues that arise with the handling of missing feedback. One way to deal with this is by sampling the missing feedback thus easily averting scalability issues. The other possibility is to smartly decompose computations into independently computable parts that can be shared through computations. We follow the latter route.

We use an Alternating Least Squares (ALS) method. In ALS only one matrix is updated at a time and all the other matrices are fixed. The optimization of the loss function is done through finding the optimal values in one feature matrix, given the others.

The two main advantages of ALS are (1) that it does not use sampling, therefore it is usually more accurate and converges faster; (2) the computations of the feature vectors -- with linear models -- are independent from each other and thus can be easily parellelized on multi-core or multiprocessor systems. The main problem with ALS is that it requires a least squares step for each feature vector computation and thus scales cubically in $K$ that makes it hard to train high factor models. Therefore we approximate the solution of the least squares problem through conjugate gradient (CG) optimization. We derive the algorithm up to efficiently computing the least squares problem where we apply CG to solve it. See earlier work \citep{cgcd_arxiv} on how to apply this learning strategy effectively.

We use the loss function from equation~\ref{eq:loss} and insert the general linear factorization model of equation~\ref{eq:model} into it with the weighting scheme described in equation~\ref{eq:simpleweight}.

Without the loss of generality, we demonstrate the calculation of $M^{(i)}$ on the $M^{(1)}$ matrix. For clearer presentation, the members of the model (equation~(\ref{eq:model})) are grouped into two based on whether a column of $M^{(1)}$ is part of them:\footnote{To avoid more complex notation, we assume that the columns of $M^{(1)}$ are the first members in the products where they are present.}
\begin{equation}\label{eq:ordered}
\begin{gathered}
    \hat{r}_{i_1,\ldots,i_{N_D}}=\\
    =\underbrace{\big(M^{(\sigma_2)}_{\pi_2}\circ\ldots\circ M^{(\sigma_{p_1})}_{\pi_{p_1}}+\ldots+M^{(\sigma_{p_{k-1}+2})}_{\pi_{p_{k-1}+2}}\circ\ldots\circ M^{(\sigma_{p_k})}_{\pi_{p_k}}\big)^T}_{\left(\mathcal{Q}_1\right)^T}M^{(1)}_{i_1}+\\[-1mm]
    +\underbrace{\big(M^{(\sigma_{p_k+1})}_{\pi_{p_k+1}}\circ\ldots\circ M^{(\sigma_{p_{k+1}})}_{\pi_{p_{k+1}}}+\ldots+M^{(\sigma_{p_{q-1}+1})}_{\pi_{p_{q-1}+1}}\circ\ldots\circ M^{(\sigma_{p_{q}})}_{\pi_{p_q}}\big)^T}_{\left(\mathcal{Q}_2\right)^T}1
\end{gathered}
\end{equation}
When recomputing $M^{(1)}$, every other matrix is fixed, thus $L$ is convex in the elements of $M^{(1)}$. The minimum is reached when $\partial L/\partial M^{(1)}$ is zero. The columns of $M^{(1)}$ can be computed separately, because the derivative is linear in them. Each column is computed similarly, therefore only the steps for $M^{(1)}_1$ (the first column of $M^{(1)}$) are shown:
\begin{equation}\label{eq:der}
\begin{gathered}
    \frac{\partial L}{\partial M^{(1)}_1}=\underbrace{-2\sum_{i_2=1,\ldots,i_{N_D}=1}^{S_2,\ldots,S_{N_D}}{r_{1,i_2,\ldots,i_{N_D}}\mathcal{W}(1,i_2,\ldots,i_{N_D})\mathcal{Q}_1}}_{\mathcal{O}}+\\[-1mm] +\underbrace{2\sum_{i_2=1,\ldots,i_{N_D}=1}^{S_2,\ldots,S_{N_D}}{w_0\hat{r}_{1,i_2,\ldots,i_{N_D}}\mathcal{Q}_1}}_{\mathcal{I}_2=\mathcal{I}+\mathcal{J}M^{(1)}_1}+ \\[-1mm] +\underbrace{2\sum_{i_2=1,\ldots,i_{N_D}=1}^{S_2,\ldots,S_{N_D}}{(\mathcal{W}(1,i_2,\ldots,i_{N_D})-w_0)\hat{r}_{1,i_2,\ldots,i_{N_D}}\mathcal{Q}_1}}_{\mathcal{I}_1=\mathcal{I}'+\mathcal{J}'M^{(1)}_1}
\end{gathered}
\end{equation}
We introduce $\mathcal{O}$, $\mathcal{I}_1=\mathcal{I}'+\mathcal{J}'M^{(1)}_1$ and $\mathcal{I}_2=\mathcal{I}+\mathcal{J}M^{(1)}_1$ to simplify further equations. $\mathcal{O}$ is the weighted sum of $\mathcal{Q}_1$ type vectors from equation~(\ref{eq:ordered}) over all possible configurations involving the first entity of the first dimension. The weights are the products of corresponding elements of the preference tensor $R$ and the value of the weighting function $\mathcal{W}$ for that setting. Due to the values of the preferences, most of the members of this sum are zero. Both $\mathcal{I}_1$ and $\mathcal{I}_2$ are the sum of a coefficient matrix multiplied by the vector we seek (i.e. $M^{(1)}_1$ in this case) and a vector. The difference is that these parts of $\mathcal{I}_2$ (i.e. $\mathcal{I}$ and $\mathcal{J}$) are the same for every column of $M^{(1)}$ (and therefore can be precomputed); while those of $\mathcal{I}_1$ (i.e. $\mathcal{I}'$ and $\mathcal{J}'$) are not.

$\mathcal{O}$, $\mathcal{I}'$ and $\mathcal{J}'$ can be computed efficiently (see section~\ref{sec:complex}), however the naive computation of $\mathcal{I}$ and $\mathcal{J}$ is expensive. Therefore we further transform $\mathcal{I}_2$. With the expansion of $\hat{r}_{1,\ldots,i_{N_D}}$ (substituting (\ref{eq:ordered}) with $i_1=1$):
\begin{equation}\label{eq:brtrf}
    \mathcal{I}_2=2w_0\sum_{i_2=1,\ldots,i_{N_D}=1}^{S_2,\ldots,S_{N_D}}{\mathcal{Q}_1(\mathcal{Q}_1)^TM^{(1)}_1+\mathcal{Q}_1(\mathcal{Q}_2)^T1}
\end{equation}
Expanding either $\mathcal{Q}_1(\mathcal{Q}_1)^T$ or $\mathcal{Q}_1(\mathcal{Q}_2)^T$ results in sums of matrix products, where the arguments are the elementwise products of multiple feature vectors:
\begin{equation}\label{eq:expr}
    \sum_{i_2=1,\ldots,i_{N_D}=1}^{S_2,\ldots,S_{N_D}}{\left(M^{(j_1)}_{i_{j_1}}\circ\ldots\circ M^{(j_m)}_{i_{j_m}}\right)\left(M^{(l_1)}_{i_{l_1}}\circ\ldots\circ M^{(l_t)}_{i_{l_t}}\right)^T}
\end{equation}
where $j_i\neq j_k$ if $i\neq k$, $l_i\neq l_k$ if $i\neq k$, $j_i\in[2\ldots n]$ and $l_k\in[2\ldots n]$. With rearranging this expression, only the following types of quantities are needed to be computed:
\begin{equation}
\begin{aligned}
\label{eq:trf}
&\text{(a)}\enskip C^{(j)}=\sum_{i=1}^{S_{j}}{M^{(j)}_{i}\left(M^{(j)}_{i}\right)^T},\\
&\text{(b)}\enskip  O^{(l)}=\sum_{i=1}^{S_{l}}{M^{(l)}_{i}},\\
&\text{(c)}\enskip S_k,
\end{aligned}
\end{equation}
where (a) $C^{(j)}\in\mathbb{R}^{K\times K}$ is the covariance matrix of the feature vectors of the $j^{\rm{th}}$ feature matrix; (b) $O^{(l)}\in\mathbb{R}^{K}$ is the sum of the feature vectors of the $l^{\rm{th}}$ feature matrix; (c) $S_k\in\mathbb{R}$ is the domain size. (\ref{eq:expr}) can be computed from (a), (b) and (c) using (1) elementwise product of $\mathbb{R}^{K\times K}$ matrices; (2) elementwise product of $\mathbb{R}^{K}$ vectors; (3) matrix product of $\mathbb{R}^{K}$ vectors; (4) matrix--scalar multiplication. Note that $S_k$ is a fix value during the training process, and $C^{(j)}$ and $O^{(j)}$ only changes after the $j^{\rm{th}}$ feature matrix is recomputed. Therefore these quantities can be precomputed and should be updated only once per epoch.

After $\mathcal{O}$, $\mathcal{I}'$, $\mathcal{J}'$, $\mathcal{I}$ and $\mathcal{J}$ from equation~(\ref{eq:der}) are computed, $\frac{\partial L}{\partial M^{(1)}_1}=0$ can be solved for $M^{(1)}_1$. Instead the least squares solver (LS), we use an approximate conjugate gradient solver to get the new value of the feature vector. Algorithm~\ref{alg:hicode} shows the high level pseudocode of the training.

\begin{algorithm}[!h]
    \caption{ALS-based learning of the general framework on implicit data}\label{alg:hicode}
    \textbf{Input:} {$T$: training data; MODEL: the description of the desired model $K$: number of features; $E$: number of epochs; $\lambda$: regularization coefficient} \newline
    \textbf{Output:} {$\{M^{(i)}\}_{i=1,\ldots, N_D}$} $K\times S_i$ sized low rank matrices \newline
    \textbf{procedure} {\hbox{\ }Train{($T$, MODEL, $K$, $E$, $\lambda$)}}
\begin{algorithmic}[1]
	\For{$i=1,\ldots,N_D$}
        \State $M^{(i)} \leftarrow $ Random $K\times S_i$ sized matrix
        \State $C^{(i)} \leftarrow \sum_{k=1}^{S_{i}}{M^{(i)}_{k}\left(M^{(i)}_{k}\right)^T}$ and $O^{(i)} \leftarrow \sum_{k=1}^{S_{i}}{M^{(i)}_{k}}$
    \EndFor
    \For{$e=1,\ldots,E$}
        \For{$i=1,\ldots,N_D$}
            \State Compute the shared parts $\mathcal{I}$ and $\mathcal{J}$
            \For{$j=1,\ldots,S_i$}
                \State Compute $\mathcal{O}$, $\mathcal{I}'$ and $\mathcal{J}'$
                \State Add regularization
                \State Solve $\frac{\partial L}{\partial M^{(i)}_j}=0$ for $M^{(i)}_j$
            \EndFor
            \State $C^{(i)} \leftarrow \sum_{k=1}^{S_{i}}{M^{(i)}_{k}\left(M^{(i)}_{k}\right)^T}$ and $O^{(i)} \leftarrow \sum_{k=1}^{S_{i}}{M^{(i)}_{k}}$
        \EndFor
    \EndFor
    \State \textbf{return} $\{M^{(i)}\}_{i=1,\ldots,N_D}$
\end{algorithmic}
\textbf{end procedure}
\end{algorithm}

Yet we neglected regularization and biases. Regularization can be done by adding a $K\times K$ sized diagonal matrix to $\mathcal{J}+\mathcal{J}'$ (i.e. to the coefficient matrix of $M^{(i)}_j$) just before computing the feature vector. The model (\ref{eq:model}) can be extended with biases by adding $\sum_{i=1}^{N_D}\sum_{j=1}^{S_i}{v_{i,j}b_{i,j}}$ to it, where $b_{i,j}$ is the bias value for the $j^{\rm{th}}$ entity of the $i^{\rm{th}}$ attribute and $v_{i,j}$ is the weight of the bias. The training of this biased model can also be done efficiently (with complexity of the non-biased $K+1$-feature model's).

\subsection{Complexity of training}\label{sec:complex}
The complexity of one epoch (i.e. computing each matrix once) is $O(N_DN^+|O|K^2+\sum_{i=1}^{N_D}{S_i}K^3)$ with a naive LS solver (see Table~\ref{tab:complex} for breakdown). This is reduced to $O(N_DN^+|O|K+\sum_{i=1}^{N_D}{S_i}K^2)$ with a carefully implemented CG solver. Since $|O|N_DN^+ \gg \sum_{i=1}^{N_D}{S_i}$ and $K$ is small ($K\in[20\ldots300]$), the first term dominates. Therefore the algorithm scales \emph{linearly} with both the number of transactions and $K$ in practice (see Section~\ref{sec:runtimes} for empirical results on running times).

\begin{table*}[!ht]
\centering
{
\caption{Complexity of computations}\label{tab:complex}
\begin{tabular}{llp{0.5\hsize}}
\toprule
\multicolumn{1}{c}{\textbf{Task}} & \multicolumn{1}{c}{\textbf{Complexity}} & \multicolumn{1}{c}{\textbf{Comments}} \\
\midrule
\multicolumn{3}{c}{\textbf{Computations required per columns of $M^{(1)}$}} \\
\midrule
$\mathcal{O}$, $\mathcal{I}'$ and $\mathcal{J}'$ & $O(N_1^+K^2|O|)$ & $N_1^+$ is the number of training events, where the value of the $A^{(1)}$ attribute is $a^{(1)}_1$, and $|O|$ is the complexity of the model (i.e. the number of vector operations to compute $\hat{r}$). This is possible due to the definition of $c$ weights and $r$ preferences, as most of the members in the sums of $\mathcal{O}$, $\mathcal{I}'$ and $\mathcal{J}'$ are in fact zeroes. \\
Solving for $M^{(1)}$ & $O(K^3)$ & Using the naive LS solver. \\
\midrule
\multicolumn{3}{c}{\pbox{10cm}{\textit{Total complexity of the above for all columns of $M^{(1)}$: $O(N^+K^2|O|+S_1K^3)$, ($N^+$ is the number of transactions)}}} \\
\midrule
\multicolumn{3}{c}{\textbf{Computations once per computing $M^{(1)}$}} \\
\midrule
Computing $\mathcal{I}$ and $\mathcal{J}$ & $O(|O|K^2)$ & Assembled from members described in equation~(\ref{eq:trf}): $C^{(j)}$ and $O^{(j)}$. These need to be recomputed when $M^{(j)}$ changes. \\
Recomputing $C^{(1)}$ and $O^{(1)}$ & $O(S_1K^2)$ & Computed after finishing the recomputation of $M^{(1)}$. \\
\midrule
\multicolumn{3}{c}{\textit{Total complexity of an epoch: $O(N_DN^+|O|K^2+\sum_{i=1}^{N_D}{S_i}K^3)$}} \\
\bottomrule
\end{tabular}
}
\end{table*}

\subsection{Special cases}
We now show that standard factorization algorithms are special cases of GFF. In standard 2D MF for implicit feedback \citep{HuICDM08}, the preference of user $u$ on item $i$ is predicted as product of user and an item features: $\hat{r}_{u,i}=1^T\left(M^{(1)}_u\circ M^{(2)}_i\right)$. iTALS -- the context-aware tensor factorization model \citep{itals_ecml} -- with 3 dimensions predicts the preference of user $u$ on item $i$ under context-state $c$ as $\hat{r}_{u,i,c}=1^T\left(M^{(1)}_u\circ M^{(2)}_i\circ M^{(3)}_c\right)$, the product of each features. Its modification -- iTALSx -- does the same by using $\hat{r}_{u,i,c}=1^T\left(M^{(1)}_u\circ M^{(2)}_i\right)+1^T\left(M^{(2)}_i\circ M^{(3)}_c\right)+1^T\left(M^{(2)}_i\circ M^{(3)}_c\right)$. If ratings are used in $R$ with $w^0(\cdot)=0$, $w^1(\cdot)=1$ and $\hat{r}_{u,i}=1^T\left(M^{(1)}_u\circ M^{(2)}_i\right)$, we get the classic ALS MF algorithm \citep{BellkorICDM07}. SVD++ \citep{Koren2008KDD} can be also derived from GFF if explicit compliant weighting and preferences are used and the items rated by the users are included through binary attributes. The model should be set accordingly to the SVD++ model. However we recommend using the extended framework (see Section~\ref{sec:gff-extended}) instead of using many binary attributes, because of increased training times.

\section{Modeling preferences under context}\label{sec:models}
In this section we demonstrate the usefulness of the GFF by examining several models therein. Recall that the preference model is an input of the framework that allows experimentation with novel models without implementing a specific algorithm. Therefore we can examine novel models and compare them and also to the traditional N-way and pairwise interaction models.

\subsection{Experimental setup}\label{sec:expsetup}
We used five genuine implicit feedback data sets to evaluate our algorithm. Three of them are public (LastFM 1K, \citep{lastfm1k}; TV1, TV2, \citep{tv1_tv2}), the other two are proprietary (Grocery, VoD). The properties of the data sets are summarized in Table~\ref{tab:data}. The column ``Multi'' shows the average multiplicity of user--item pairs in the training events.\footnote{This value is 1.0 at TV1 and TV2. This is possibly due to preprocessing by the original authors that removed duplicate events.} The train--test splits are time-based: the first event in the test set is after the last event of the training set. The length of the test period was selected to be at least one day, and depends on the domain and the frequency of events. We used the artists as items in LastFM.

\begin{table}
\centering
\caption{Main properties of the data sets}\label{tab:data}
\medskip
{\small
\begin{tabular}{@{}l@{\hskip2mm}l@{\hskip4mm}r@{\hskip2mm}r@{\hskip2mm}r@{\hskip2mm}r@{\hskip4mm}r@{\hskip2mm}l@{}}
\toprule
\multirow{2}{*}{\textbf{Dataset}} & \multirow{2}{*}{\textbf{Domain}}& \multicolumn{4}{c}{\textbf{Training set}}& \multicolumn{2}{c}{\textbf{Test set}} \\
&& \textbf{\#Users}& \textbf{\#Items}& \textbf{\#Events}& \textbf{Multi}& \textbf{\#Events} & \textbf{Length}\\
\midrule
Grocery & E-grocery & 24947 & 16883 & 6238269 & 3.0279 & 56449 & 1 month \\
TV1 & IPTV & 70771 & 773 & 544947 & 1.0000 & 12296 & 1 week \\
TV2 & IPTV & 449684 & 3398 & 2528215 & 1.0000 & 21866 & 1 day \\
LastFM & Music & 992 & 174091 & 18908597 & 21.2715 & 17941 & 1 day \\
VoD & IPTV/VoD & 480016 & 46745 & 22515406 & 1.2135 & 1084297 & 1 day \\
\bottomrule
\end{tabular}}
\end{table}

We focus on topN recommendations. For a given user--context configuration setting all items are ranked by their predicted preference ($\hat{r}$). The first $N=20$ items of the list are used for recommendations.

Our primary evaluation metric is recall@20,\footnote{We also measured recall@10 and recall@5 (not shown); the relation between different models are the same.} defined as the ratio of relevant recommended items and relevant items. The reason for using recall@N is twofold: (1) we found that in live recommender systems recall usually correlates well with click-through rate (CTR), that is, an important online metric for recommendation success. (2) As described in \citep{itals_ecml}, recall@20 is a good proxy of estimating recommendation accuracy offline for real-world applications; similar finding is available in \citep{Liu:2012EBR}.\footnote{If we have no highlighted items in the recommendations (i.e. all recommended items are equal), then it makes sense to disregard the order of the recommended items. Whether this is true is determined by both the interface and the recommendation logic. For example, if we want to show more items or more diverse itemset to a user during a session while still giving relevant recommendations, we can randomize the top $N$ recommendation and recommend the first $K$ of this random order. This way we can overcome showing users the same $K$ items multiple times and have a higher chance for clicking. The goal of the system is to recommend items that the user likes. The @20 comes from a very average setting of recommending 5 items (from a randomized pool of top 20 items) per page and the user having 4--6 page views in a session. Of course these numbers are highly varied in different applications, but we still think that this is a realistic proxy for a real recommender as it can get.} Note also that recall is event based, while ranking based metrics like MAP and NDCG are query based. The inclusion of context changes the query set of the test data, therefore the comparison by query based metrics is unfair.


The hyperparameters of the algorithm, such as regularization coefficients, were optimized on a part of the training data (validation set). Then the algorithm was trained on the whole training data (including the validation set) and recall was measured on the test set. The number of epochs was set to 10 in all cases, because (1) we found that methods converge fairly well in at most 10 epochs; (2) the time of the training should be also considered and 10 epochs is usually a good trade-off between time and accuracy in practical settings. The number of features was generally set to $K=80$ that is considered to be between low and high factor models and is usually a good setting in practice. However we conduct an experiment regarding the effect of $K$ on the performance of selected models.

\subsubsection{A context-aware problem}
Our aim is to apply GFF to the implicit feedback based context-aware recommendation problem and find models that generally perform well. The area of context-aware problems is wide, as any additional information to the user--item interaction can be considered as context. In compliance with SA-MDM, we assume that the context dimensions are event contexts, meaning that their value is not determined solely by the user or the item; rather it is bound to the transaction. E.g. the time of the transaction is an event context, while the genres of the item is not.

Here we choose a general CA setup and use the \emph{time} and \emph{the order of the transactions} to derive context variables that are relevant and thus help improving recommendation accuracy. Implicit feedback data does not typically contain many other event context variables: some contexts, like \emph{mood}, require to be explicitly stated, while others, like \emph{location, device}, are specific to domains. Thus, \emph{seasonality} and \emph{sequentiality} are applied as contexts of the transaction. Therefore the problem we use as an example consists of four dimensions, a transaction is a 4-tuple that contains (1) the user, (2) the item, (3) the time band (based on the timestamp), (4) and the previously consumed item by the same user.

\textbf{Seasonality:} Many application areas of recommender systems exhibit the seasonality effect, because periodicity can be observed in many human activities. Therefore seasonal data is an obvious choice for context \citep{LiuCAMRA10}. First we have to define the length of the season. Within a season we do not expect repetitions in the aggregated behavior of users, but we expect that at the same time offset in different seasons, the aggregated behavior of the users will be similar. The length of the season depends on the data. Once we have this, within seasons we need to create \emph{time bands} (bins) that are the possible context-states. Time bands specify the time resolution of a season, which is also data dependent. We can create time bands with equal or different length. In the final step, events are assigned to time bands according to their time stamp.

For \emph{Grocery} we defined a week as the season and the days of the week as the time bands. The argument here is that people usually do shopping on weekly or biweekly basis and that shopping habits differ on weekends and weekdays. One day was used as season for the other four data sets with 4 hour intervals. We note that one can optimize the lengths and distribution of time bands but this is beyond the scope of the current paper.

\textbf{Sequentiality:} In some domains, like movies or music, users consume similar items. In other domains, like electronic gadgets or e-commerce in general, they avoid items similar to what they already consumed and look for complementary products. Sequential patterns can be observed on both domain types. Sequentiality was introduced in \citep{itals_ecml} and uses the previously consumed item by the user as a context for the actual item. This information helps in the characterizations of repetitiveness related usage patterns and sequential consumption behavior.

During evaluation we fix the sequential context to the item that was targeted by the last transaction of the user in the training set. Thus we do not use information from the test data during the evaluation. The other way (i.e. constantly update the context value based on test events) would be valid as well and would result in better results. Because the test data spans over a short period of time that generally contains a few purchasing sessions for the users, preferences thus can be accurately predicted also from this information.

\subsection{Preference models}
First, we introduce a highly simplified notation for preference models. We denote the four dimensions by $U$, $I$, $S$ and $Q$ for users, items, seasonality and sequentiality respectively. The models consists of selected interactions between selected dimensions. An interaction is denoted by putting the dimensions after one another. E.g. $UI$ is the user--item interaction, $USI$ is the user--item--seasonality interaction and so on. A model usually contains more than one interaction. Table~\ref{tab:expl} shows examples of this notation.

\begin{table}
\centering
\caption{Examples of the simplified notation system}\label{tab:expl}
\medskip
{\small
\begin{tabular}{ll}
\toprule
\multirow{2}{*}{$UI$} & Vanilla MF model (user--item interactions): \\
&  $\hat{r}_{u,i}=1^T\left(M^{(U)}_u\circ M^{(I)}_i\right)$ \\
\midrule
\multirow{2}{*}{$USQI$} & N-way model with all 4 dimensions (tensor factorization): \\
& $\hat{r}_{u,i,s,q}=1^T\left(M^{(U)}_u\circ M^{(I)}_i\circ M^{(S)}_s\circ M^{(Q)}_q\right)$ \\
\midrule
\multirow{2}{*}{$UI+US+IS$} & Pairwise interaction model with 3 dimensions (U, I, S): \\
& $\hat{r}_{u,i,s}=1^T\left(M^{(U)}_u\circ M^{(I)}_i+M^{(U)}_u\circ M^{(S)}_s+M^{(I)}_i\circ M^{(S)}_s\right)$ \\
\bottomrule
\end{tabular}}
\end{table}

There are 11 different possible interactions with 4 dimensions therefore the number of possible preference models is $2^{11}-1=2047$. Removing the ones that do not contain $U$ or $I$, we still get 2018 potential models. In the field of context-aware recommendations state-of-the-art factorization methods use two models. The pairwise interaction model ($UI+US+IS+UQ+IQ+SQ$ with all 4 dimensions) (\citet{RendleWSDM2010,rendle2012factorization,italsx_infocomm}) assumes pairwise interaction between each pair of dimensions. On the other hand the N-way model ($UISQ$ with all 4 dimensions) (\citet{itals_ecml,Shi_tfmap}) assumes that the preferences can be best described by the joint interaction of all dimensions. \citep{rendle2012factorization} also mentions the generalization of the pairwise interaction model, coined d-way interaction model (e.g. the 3-way interaction model: $UI+US+IS+UQ+IQ+SQ+USI+UQI+USQ+ISQ$ with all 4 dimensions). This model includes all interactions between subsets of dimensions up to $d$ size. The authors argue that such a model is slow to train and usually does not result in more accurate recommendations.

We approach the preference modeling from the perspective of the context-aware recommendation task. In this setting the users initiate transactions with the items. Additional variables (context) may or may not influence user behavior, therefore not all possible interactions should be considered for preference modeling. We focus on the ones where either the user, the item or both interact with a context. Interactions where contexts interact with each other are disregarded (see Section~\ref{sec:independence} for additional justification), except for $SQ$ that we only keep for compatibility's sake with the pairwise model. Therefore we get to the followings:
\begin{itemize}[noitemsep]
\item \textbf{$\boldsymbol{UI}$:} Interaction between users and items, the classic CF model.
\item \textbf{$\boldsymbol{USI}$, $\boldsymbol{UQI}$, $\boldsymbol{USQI}$:} The context value dependent reweighting of the user--item relation, i.e. the context influences how the users interact with items. More context dimensions can be used for reweighting. But the more we use, the more sensitive it becomes to noise and more latent features are required for filtering this out \citep{italsx_infocomm}.
\item \textbf{$\boldsymbol{US}$, $\boldsymbol{UQ}$:} The user--context interaction produces a context dependent user bias that does not play role during the ranking but has noise filtering properties during training. We allow only one context in these interactions, because additional contexts would assume that different context dimensions interact somehow.
\item \textbf{$\boldsymbol{IS}$, $\boldsymbol{IQ}$:} The item--context interaction results in a context dependent item bias that helps in ranking as well as in learning. Only one context is allowed in these interactions.
\item \textbf{$\boldsymbol{SQ}$:} Interactions between the two context dimensions. Required for the traditional pairwise model.
\end{itemize}

\begin{figure*}[!ht]
\centering
\includegraphics[width=\textwidth]{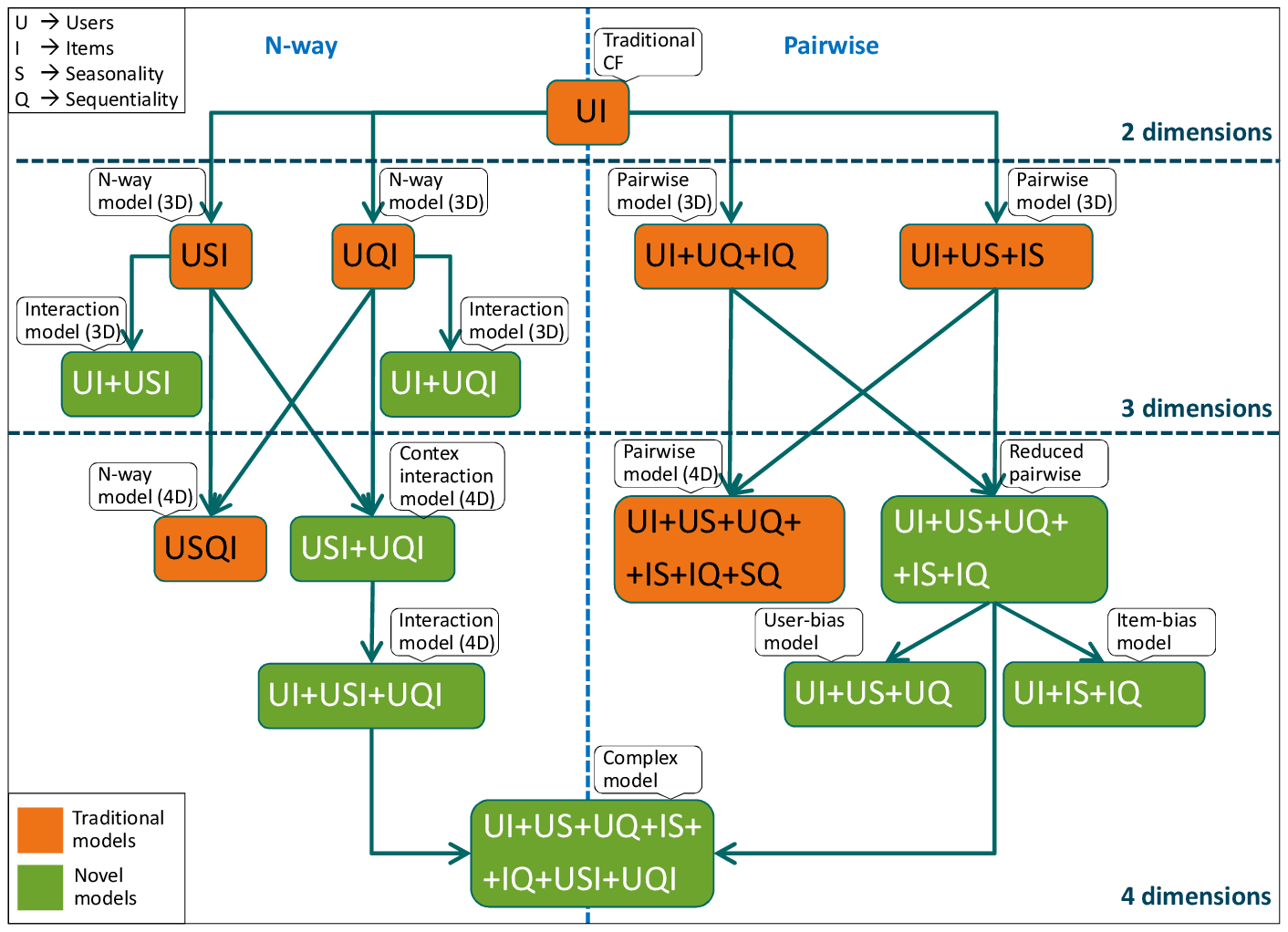}
\caption{Hierarchy of the models}
\label{fig:models}
\end{figure*}

The models we used are depicted on Figure~\ref{fig:models}. Models on the right side follow the pairwise interaction scheme, while models on the left are of the N-way flavor. \emph{Traditional models} -- that were used also earlier -- are indicated with orange background and black text and \emph{novel models} are with green background and white text. The models are sorted to layers based on the dimensions used. In 2D there is only the classical $UI$ model of CF. With the inclusion of one context dimension (either $S$ or $Q$) the N-way and the pairwise philosophy of preference modeling diverges. There are only a few novel models with three dimensions and we only selected those that we coined \emph{interaction model}. Things get interesting with all four dimensions where one can create many novel models. We selected the following novel models for experimentation:
\begin{itemize}[noitemsep]
\item \textbf{Interaction model ($\boldsymbol{UI+USI+UQI}$):} This model is the composite of the base behavior of the users ($UI$) and their context-influenced modification of this behavior ($USI$ and $UQI$). This model assumes that the preferences of the users can be divided into context independent and dependent parts. In the latter the user--item relation is reweighted by a context dependent weight vector. $USQI$ is not included due to the noisiness of reweighting by more than one weight vector simultaneously.
\item \textbf{Context interaction model ($\boldsymbol{USI+UQI}$):} Preferences in this model are modeled by solely context dependent parts, i.e. it assumes that user--item interactions strongly depend on the context and this dependency affects the whole interaction rather than solely the items or users.
\item \textbf{Reduced pairwise model ($\boldsymbol{UI+US+IS+UQ+IQ}$):} This model is a minor variation of the traditional pairwise model with the exclusion of the interaction between context dimensions ($SQ$). The interaction with context is done separately by users and items, i.e. it does not affect the whole user--item relation.
\item \textbf{User bias model ($\boldsymbol{UI+US+UQ}$):} Here it is assumed that only the user interacts with the other dimensions. This results in a model where the user--item relation is supported by context dependent user biases. Note that during recommendation the user biases are constant, thus do not affect the ranking. However they might filter out some context related noise during training.
\item \textbf{Item bias model ($\boldsymbol{UI+IS+IQ}$):} This model assumes that the effect of context can described by context dependent item biases (e.g. items are popular under certain conditions). The item biases affect the ranking as well as filter context related noise during training.
\item \textbf{A complex model ($\boldsymbol{UI+US+IS+UQ+IQ+USI+UQI}$):} This model is the composite of the reduced pairwise and the interaction model. It can be also treated as a reduced 3-way interaction model from which the context-context interactions are omitted.
\end{itemize}

Note, that we restricted our model space to those where exactly one feature matrix belongs to each dimension. In GFF it is possible to use several set of features for selected dimensions. By doing so it is possible to decouple the modeling of different effects from each other. For example user and item interaction with a certain context dimension can be modeled separately by using two sets of feature for the context dimension. This is a far reaching research direction that is out of the scope of this paper, but nonetheless made available by GFF.

\subsection{Results}

\begin{table}[!h]
\centering
{
\caption{Recall@20 values for different models within the framework. Differences are statistically significant at $p=0.05$. Traditional models are with gray background. Best results are typeset bold.}\label{tab:results}
\begin{tabular}{lrrrrr}
\toprule
\textbf{Model} & \textbf{Grocery} & \textbf{TV1} & \textbf{TV2} & \textbf{LastFM} & \textbf{VoD} \\
\midrule
\pbox{4.7cm}{$USI+UQI$ \\ (context interaction model)} & 0.1504 & \textbf{0.1551} & 0.2916 & 0.1984 & 0.1493 \\[0.25cm]
\pbox{4.7cm}{$UI+USI+UQI$ \\ (interaction model)} & \textbf{0.1669} & 0.1482 & \textbf{0.3027} & \textbf{0.2142} & \textbf{0.1509} \\[0.25cm]
\rowcolor{LGray} \pbox{4.7cm}{$USQI$ \\ (N-way model)} & 0.1390 & 0.1315 & 0.2009 & 0.1906 & 0.1268 \\[0.25cm]
\pbox{4.7cm}{$UI+US+IS+UQ+IQ$ \\ (reduced pairwise model)} & 0.1390 & 0.1352 & 0.2388 & 0.1884 & 0.0569 \\[0.25cm]
\pbox{4.7cm}{$UI+US+UQ$ \\ (user bias model)} & 0.1619 & 0.0903 & 0.1399 & 0.1993 & 0.0335 \\[0.25cm]
\pbox{4.7cm}{$UI+IS+IQ$ \\ (item bias model)} & 0.1364 & 0.1266 & 0.2819 & 0.1871 & 0.1084 \\[0.25cm]
\rowcolor{LGray} \pbox{4.7cm}{$UI+US+IS+UQ+IQ+SQ$ \\ (pairwise interaction model)} & 0.1388 & 0.1344 & 0.2323 & 0.1873 & 0.0497 \\[0.25cm]
\pbox{4.7cm}{$UI+US+IS+UQ+IQ+USI+UQI$ \\ (complex model example)} & 0.1389 & 0.1352 & 0.2427 & 0.1866 & 0.0558 \\
\bottomrule
\end{tabular}
}
\end{table}

Table~\ref{tab:results} shows the accuracy in terms of recall@20 of two traditional models and the six novel models we just introduced.

There exists a novel model with all five datasets that performs better than the both traditional models. 4 out of 5 cases the interaction model ($UI+USI+UQI$) is the best and it is the second best in the remaining one case. Thus this model is not only intuitively sound but also performs well that underpins its assumptions on preference modeling. The context interaction model ($USI+UQI$) come second in 3, and third in 2 cases. Interestingly the user bias model ($UI+US+UQ$) is the second best in 2 out 5 cases while worst one in the other 3 cases. This can be explained by the differences between the repetitiveness of the datasets. Highly repetitive datasets affected more heavily by sequentiality and benefit from the noise filtering property of the $UQ$ member.As sequentiality is more closely related to user behavior than to the items, $UQ$ is much more effective than $IQ$.

The reduced pairwise model is better than the full pairwise interaction model in all cases, however the difference is negligible in 3 out of 5 cases. But the difference is $\sim14.44\%$ by VoD and $\sim2.8\%$ by TV2 dataset. Finally, note that the complex model generally does not improve over the reduced pairwise model considerably and is always worse than the interaction and the context interaction models. Three way interactions contribute to the score in a lesser way, because features being generally less than 1 and thus three way products give smaller values. This causes the context dependent biases to be more prominent initially, thus the features are set accordingly to optimize the bias values. This confirms the observations by \citep{rendle2012factorization} finding the d-way interaction model no more useful than the pairwise interaction model. However this problem might be tackled by using two sets of features for $S$ and $Q$, separately for the three way interactions and context dependent biases.

\begin{table}[!h]
\centering
{
\caption{Improvements over traditional models}\label{tab:summary}
\begin{tabular}{llcp{4cm}}
\toprule
\textbf{Dataset} & \textbf{Best model} & \textbf{Improvement} &\hfill \textbf{Models better than traditional ones (out of 6)}\\
\midrule
Grocery & $UI+USI+UQI$ & +20.14\% & \multicolumn{1}{c}{3} \\
TV1 & $USI+UQI$ & +15.37\% & \multicolumn{1}{c}{2} \\
TV2 & $UI+USI+UQI$ & +30.30\% & \multicolumn{1}{c}{5} \\
LastFM & $UI+USI+UQI$ & +12.40\% & \multicolumn{1}{c}{3} \\
VoD & $UI+USI+UQI$ & +19.02\% & \multicolumn{1}{c}{2} \\
\bottomrule
\end{tabular}
}
\end{table}

Table~\ref{tab:summary} summarizes the improvements by novel models over the traditional ones. The best novel model (interaction model in 4/5 and context interaction model 1/5) outperforms the best traditional model by 12--30\% in terms of recall@20. This difference is significant. Besides, there are several novel models for each dataset that outperform the traditional models by more than $2\%$. These include the context interaction model and models specifically good for the data (e.g. the user bias model for Grocery and LastFM).

\begin{figure*}[!ht]
\centering
\includegraphics[width=\textwidth]{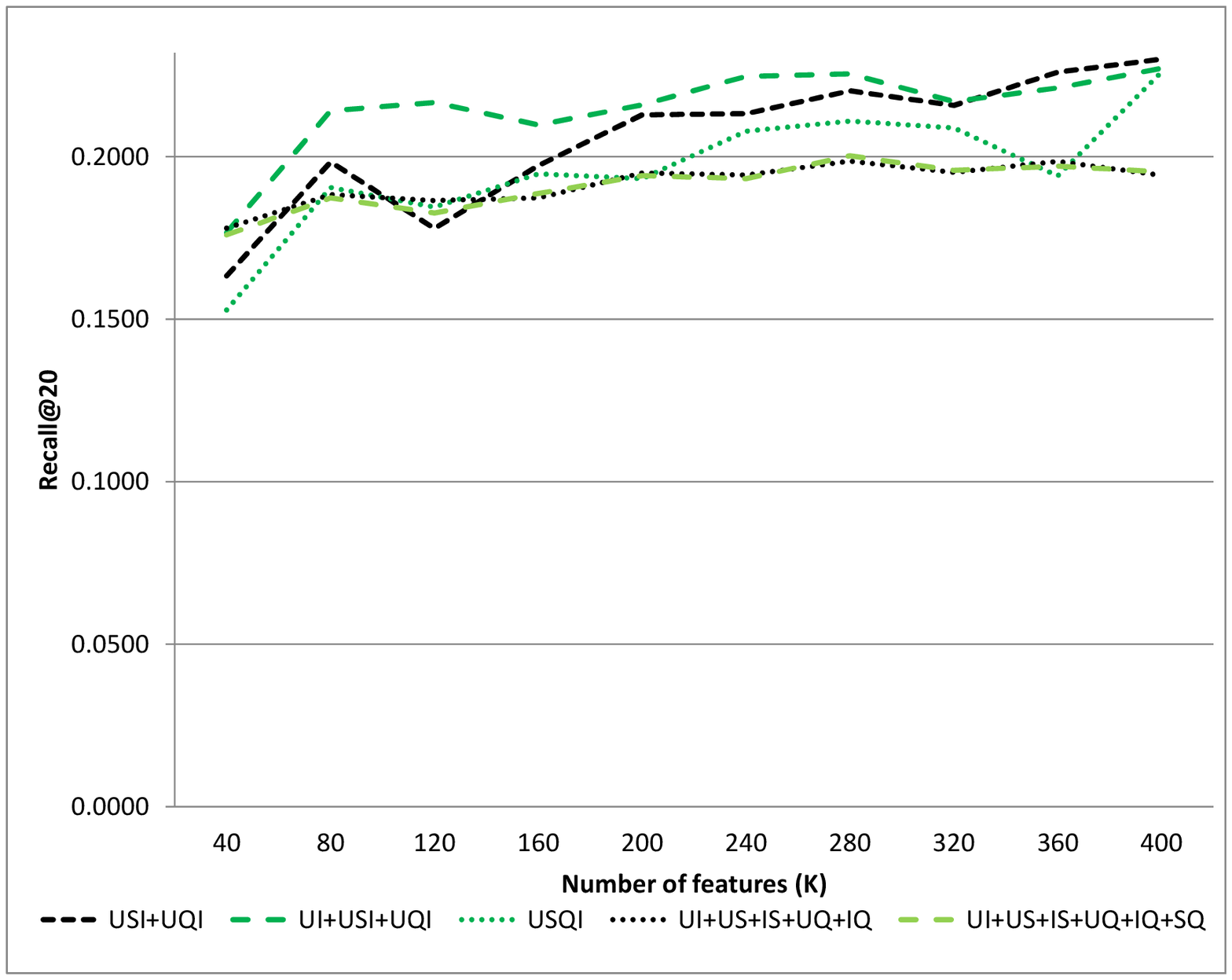}
\caption{Model accuracy versus number of factors.}
\label{fig:numfactors}
\end{figure*}

So far the number of features was fixed at $K=80$. However this parameter can significantly affect the relation of models. In earlier work \citep{italsx_infocomm} we compared the pairwise and the N-way model on two 3 dimensional context-aware problems. We found that pairwise models perform better with lower number of factors, but N-way models improve more rapidly as $K$ increases. This is due to low factor models blurring different aspects of the entities together thus making the reweighting of the N-way model more difficult if not impossible.

Figure~\ref{fig:numfactors} depicts recall@20 for different values of $K$ ranging from 40 to 400. Five models were selected for this experiment: the well performing interaction ($UI+USI+UQI$) and context interaction ($USI+UQI$) model; the traditional N-way ($USQI$) and pairwise ($UI+US+IS+UQ+IQ+SQ$) model; and the reduced pairwise model ($UI+US+IS+UQ+IQ$). The results are presented on the LastFM dataset. At $K=40$ $USI+UQI$ and $USQI$ are clearly worse than the other models and the reduced pairwise model is even slightly better than $UI+USI+UQI$ and the pairwise model. By $K=400$ the context interaction model is leading slightly (within $2\%$) compared to the interaction and the traditional N-way models. The pairwise and reduced pairwise models on the other hand lag behind by more than $15\%$. We can observe that as $K$ increases, the accuracy of models with members of higher order of interactions increase more rapidly. The N-way model improves the fastest and would probably outperform other models if the $K$ is sufficiently high. On the other hand, larger $K$ values (at or beyond 400) require longer training time and even more importantly, comes with longer recommendation times, therefore their practical use is limited.

It is also worth noting that $UI+USI+UQI$ performs more stable than $USI+UQI$ or the N-way model. This is due to the $UI$ part (i.e. the context independent user--item relation) stabilizing the prediction. We conclude that the interaction model ($UI+USI+UQI$) performs well not just for $K=80$ but for practically important $K$ values in general.

\subsection{Context-context interactions}\label{sec:independence}
We discussed that interactions between context dimensions should be excluded from the model. Comparing the pairwise and reduced pairwise model showed that modeling such interactions does not increase accuracy and sometimes even degrades it. From a recommendation task perspective context dimensions never interact with each other. They can influence the users' behavior (also via their active/inactive status), and through the users they affect the consumption pattern of items as well. One could argue that other context dimensions are also affected in a similar way. However recall that not all dimensions are equal and the main focus in recommendation is to recommend items to the users (under different contexts). Even if certain context dimensions correlate there is no direct interaction between them.

We also argue that context dimensions should be independent from each other. The context-aware recommendation task becomes harder \citep{nguyen2014gaussian} and slower \citep{Rendle2013VLDB} as the number of dimensions increase. Therefore context dimensions should ideally capture different aspects of the data rather than describing the same or highly correlated characteristics in different ways.

The context dimensions of the example setting ($S$ and $Q$) are fairly independent from each other. To quantify the independence of two context dimensions $C^{(1)}$ and $C^{(2)}$, the following probability distributions can be approximated from the training data: $P(C^{(1)})=\left\{P(C^{(1)}=c^{(1)}_i)\right\}$ and $P_j(C^{(1)})=\left\{P(C^{(1)}=c^{(1)}_i|C^{(2)}=c^{(2)}_j)\right\}$. The average Kullback--Leibler divergence between $P(C^{(1)})$ and $P_j(C^{(1)}|C^{(2)})$ for all $j$ can be then computed. Small average KL divergence means that $P(C^{(1)})$ can be used in the place of $P_j(C^{(1)}|C^{(2)})$ distributions. In other words knowing the state in $C^{(2)}$ gives us low information on the state in $C^{(1)}$.

This experiment was executed with $C^{(1)}=C^{(2)}=S$ (totally dependent context dimensions); $C^{(1)}=S$, $C^{(2)}=Q$ (sequentiality's information on seasonality); $C^{(1)}=Q$, $C^{(2)}=S$ (seasonality's information on sequentiality); $C^{(1)}=S$, $C^{(2)}=S'$; $C^{(1)}=S'$, $C^{(2)}=S$, where $S'$ is seasonality with the same season as $S$, but uses different time bands. The results are shown in table~\ref{tab:kl}. It is obvious that seasonality has little information on sequentiality and vice versa, therefore these context dimensions hardly correlate. This explains why the full pairwise model performs worse than the reduced pairwise model.

\begin{table*}[!ht]
\centering
{
\caption{Average KL divergences from $P(C^{(1)})$ to $P_j(C^{(1)}|C^{(2)})$}\label{tab:kl}
\begin{tabular}{llllll}
\toprule
\multirow{3}{*}{\textbf{Data set}} & \multicolumn{5}{c}{\textbf{Average $D_{KL}\left(P_j(C^{(1)}|C^{(2)})||P(C^{(1)})\right)$}} \\
& $C^{(1)}=S$ & $C^{(1)}=S$ & $C^{(1)}=S$ & $C^{(1)}=Q$ & $C^{(1)}=S'$ \\
& $C^{(2)}=S$ & $C^{(2)}=Q$ & $C^{(2)}=S'$ & $C^{(2)}=S$ & $C^{(2)}=S$ \\
\midrule
Grocery & 3.2574 & 0.0696 & 2.2997 & 0.0695 & 2.5238 \\
TV1 & 3.1032 & 0.0189 & 1.7235 & 0.0171 & 1.5203 \\
TV2 & 2.8132 & 0.0811 & 2.7979 & 0.0947 & 2.7707 \\
LastFM & 2.6376 & 0.0030 & 2.6162 & 0.0976 & 2.5618 \\
VoD & 2.6300 & 0.0262 & 1.8024 & 0.0547 & 2.1650 \\
\bottomrule
\end{tabular}
}
\end{table*}

\subsection{Training time}\label{sec:runtimes}

\begin{figure*}[!ht]
\centering
\includegraphics[width=\textwidth]{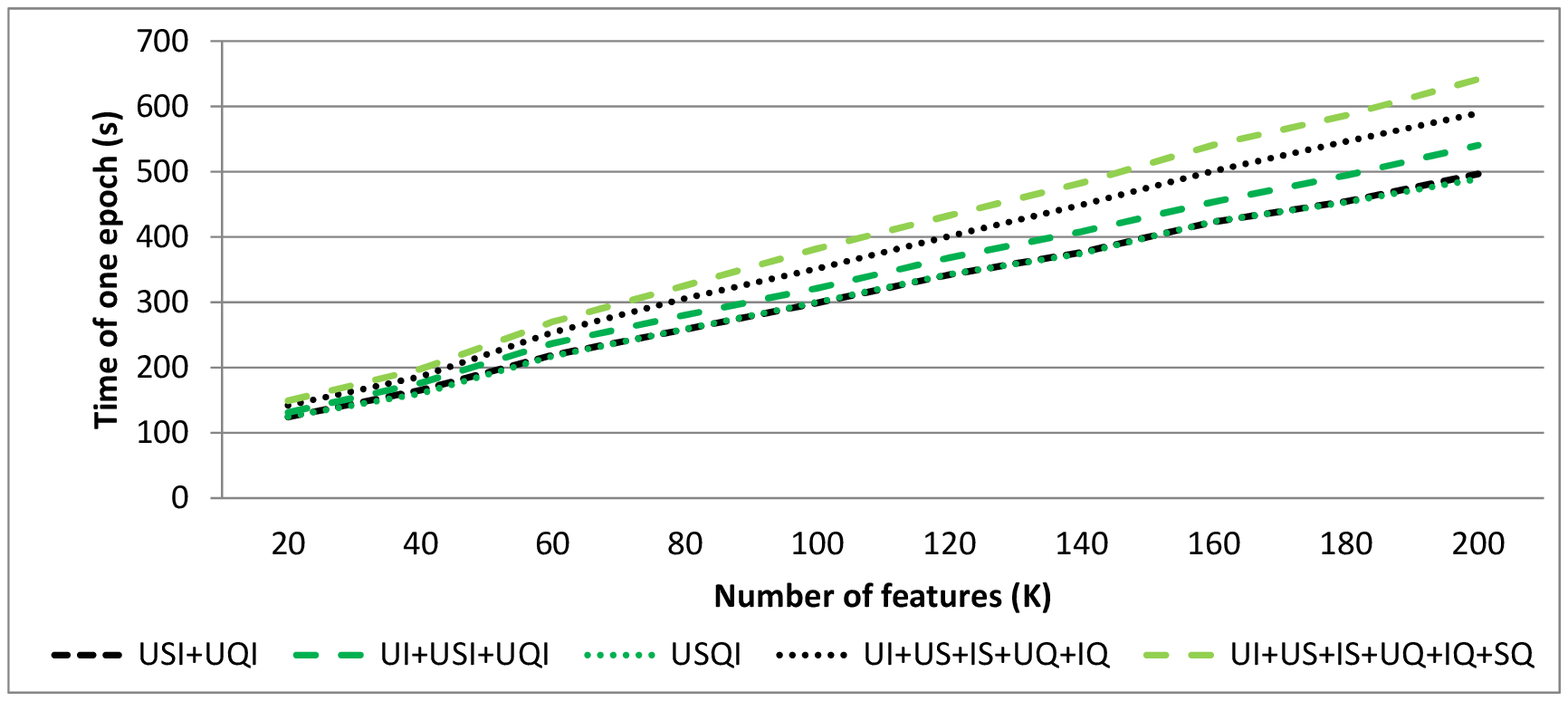}
\caption{Training times of models}
\label{fig:runtimes}
\end{figure*}

Figure~\ref{fig:runtimes} shows the time of one epoch (i.e. computing each feature matrix once) for selected models for different values of $K$ on the VoD dataset. The experiments were carried out using a single core of a multi-core CPU machine. Note that the computation can be easily parallelized therefore these training times can be greatly reduced in practice. As stated in Section~\ref{sec:complex}, the running time scales linearly with the number of features for $K$ in the practically useful range. There is a difference between the actual time of training for different models as it also depends on the complexity of the model. The complexity of the model is the number of operations required to compute the preference model. If the set of dimensions is fixed, the scaling in the model complexity is linear. In accordance with this the N-way model is the fastest and the pairwise model is the slowest from the selected ones. Also note that modeling the useless $SQ$ interaction also slows down the training.

\section{Comparison with state-of-the-art algorithms}\label{sec:comp-other}
In this section we compare GFF with other methods. The qualitative comparison focuses on pointing out key differences between GFF and other factorization algorithms. Although the main advantage of GFF is not necessarily that it can outperform other methods, but rather its flexibility (regarding the model and weighting); we also include a quantitative comparison with widely accepted algorithms such as Factorization Machines (FM) \citep{rendle2012factorization} and Bayesian Personalized Ranking (BPR) \citep{bpr}.

\subsection{Qualitative comparison with factorization methods}

\subsubsection{Factorization Machines}
Rendle et.\ al proposed factorization machines (FM; \citep{rendle2012factorization}) as a general factorization method. It is for rating prediction (explicit flavor). Implicit feedback problem can be tackled through subsampling negative feedback. Each rating is associated with different attributes, for example the user who rated, the item that was rated, the context of the rating, metadata of the item, etc. The preference model is a full (weighted) pairwise model: the prediction score is given by the sum of pairwise interaction scores between every pair of dimensions.\footnote{We rephrased here the feature matrix based introduction of the original paper.} The weight of a certain interaction is determined by the weight of the two corresponding attributes; this is an input of the algorithm. It builds on the SA-MDM datamodel just like basic GFF, therefore it handles composite dimensions through binary variables as dimensions. This solution has two drawbacks: (1) it significantly increases the training time; (2) and a lot of unnecessary interactions are modeled between these binary attributes. The authors proposed a partitioning method to overcome this problem in \citep{Rendle2013VLDB}, which basically results in excluding certain interactions from the pairwise model. The latent feature vectors can be learnt by several learning methods: stochastic gradient descent (SGD), coordinate descent\footnote{A certain version of ALS, which optimizes for one parameter at a time.}, adaptive SGD and a Bayesian inference using Markov Chain Monte Carlo (MCMC). The latter is advised as the best one of the four. The implementation of FM is available in libFM.\footnote{\url{http://libfm.org}}

The key differences between GFF and FM are as follows: (1) FM uses a subset of all possible pairwise interactions between dimensions, while GFF can use arbitrary linear preference model. (2) FM handles implicit feedback through subsampling the missing (negative) feedback and is mainly an explicit method. GFF smartly decomposes computations therefore does not need to sample implicit feedback and with the proper weighting it can either be an implicit or an explicit method. (3) Both basic GFF and FM builds on SA-MDM, however the extended GFF (see Section~\ref{sec:gff-extended}) is fully compliant with the more extensive MDM. (4) The optimization strategy of the two methods differ.

\subsubsection{SVDFeature}
Chen et.\ al proposed another framework, coined SVDFeature, that uses a subset of the FM model \citep{chen2012feature,Chen12JMLR}. Basically it assigns each attribute either to the user or to the item as a property. A feature vector is defined for each property (including the item and the user itself), and the feature vector of the item (or user) is the weighted sum of the feature vectors of its properties. The rating is predicted by the scalar product of these aggregated feature vectors. In other words, it uses a partial pairwise model that only keeps the interactions between item and user attributes. The authors claim that doing so the training time decreases drastically compared to that of FM, and the interactions dropped are mostly useless (such as interactions between metadata terms of the items). Our experiments also show that leaving out useless interactions results in more accurate models. SVDFeature can incorporate either explicit or implicit feedback as it uses a ranking loss function. The model is learned using SGD. The datamodel is a dimension restricted MDM with only 2 dimensions, one for users and one for items.

The key differences between GFF and SVDFeature are as follows: (1) SVDFeature uses a fixed model, GFF takes the preference model as an input. (2) The methods use different data models, although the data model of SVDFeature is a special case of the data model of the extended GFF. (3) SVDFeature uses (pairwise) ranking loss, GFF uses pointwise ranking loss. (4) The optimization strategies differ.

Due to the incompatibility between the data model of the basic GFF and SVDFeature, and the perception of context -- i.e. a context should be assigned to either the items or the users -- no direct quantitative comparison is possible.

\subsubsection{Other implicit context-aware factorization algorithms}
iTALS \citep{itals_ecml} and iTALSx \citep{italsx_infocomm} are general factorization algorithms that use the N-way and pairwise models respectively. The key difference to GFF is that GFF does not use a fixed model. By setting the appropriate preference model, iTALS and iTALSx are special cases of GFF.

TFMAP \citep{Shi_tfmap} is a tensor factorization algorithm for three dimensional context-aware problems that minimizes a listwise ranking loss function with SGD on a fixed 3-way model. GFF is much more flexible as TFMAP restricts not just the model class, but also the number of dimensions. The loss function and the optimization strategy of the two methods also differ.

\subsection{Quantitative comparison}
Although we argue that the main novelty and the importance of GFF is allowing experimentation with novel models without requiring specific implementations, a quantitative comparison to Factorization Machines (state-of-the-art in context-aware factorization) and to Bayesian Personalized Ranking (state-of-the-art in handling implicit feedback) is included in this section. Both FM and BPR require the missing (negative) feedback to be sampled. We followed the steps of \citet{nguyen2014gaussian} and sampled a negative example to each positive example by replacing the item of the positive example with an item that has never occurred in the training set with the same user and context values. For FM we assigned ratings 1 and 0 to positive and negative feedbacks, respectively.

FM was trained using MCMC that is encouraged by the authors of the method. The number of factors was set to $K=80$ and the number of iterations was set to $10$, because of practical requirements in the training time. Also, the method converged fairly well in $10$ epochs. There were no additional hyperparameters to be optimized by FM.

The number of features and iterations was set to $K=80$ and $10$ respectively for BPR as well. The regularization coefficients and learning rate were optimized in the same way we optimized hyperparameters for GFF.

\begin{table}[!h]
\centering
{
\caption{Comparison of GFF models to LibFM and BPR}\label{tab:comp}
\begin{tabular}{lccccc}
\toprule
\multirow{2}{*}{Dataset} & \multicolumn{3}{c}{GFF} & \multirow{2}{*}{LibFM} & \multirow{2}{*}{BPR} \\
& N-way & Pairwise & Best non-traditional && \\
\midrule
Grocery & 0.1390 & 0.1388 & \textbf{0.1669} & 0.0912 & 0.1412 \\
TV1 & 0.1315 & 0.1344 & 0.1551 & \textbf{0.1683} & 0.1365 \\
TV2 & 0.2009 & 0.2323 & 0.3027 & \textbf{0.3081} & 0.1957 \\
LastFM & 0.1906 & 0.1873 & \textbf{0.2142} & 0.0652 & 0.2002 \\
VoD & 0.1268 & 0.0497 & \textbf{0.1509} & 0.1151 & 0.0539 \\
\bottomrule
\end{tabular}
}
\end{table}

Table~\ref{tab:comp} shows the results (recall@20). For GFF, the pairwise, the N-way and the best non-traditional model (either interaction or context interaction model) was included. GFF outperforms FM in 3 out of 5 cases, performs very similarly in 1 case and underperforms in 1 case. GFF outperforms BPR in all cases.

\begin{figure*}[!h]
\centering
\includegraphics[width=\textwidth]{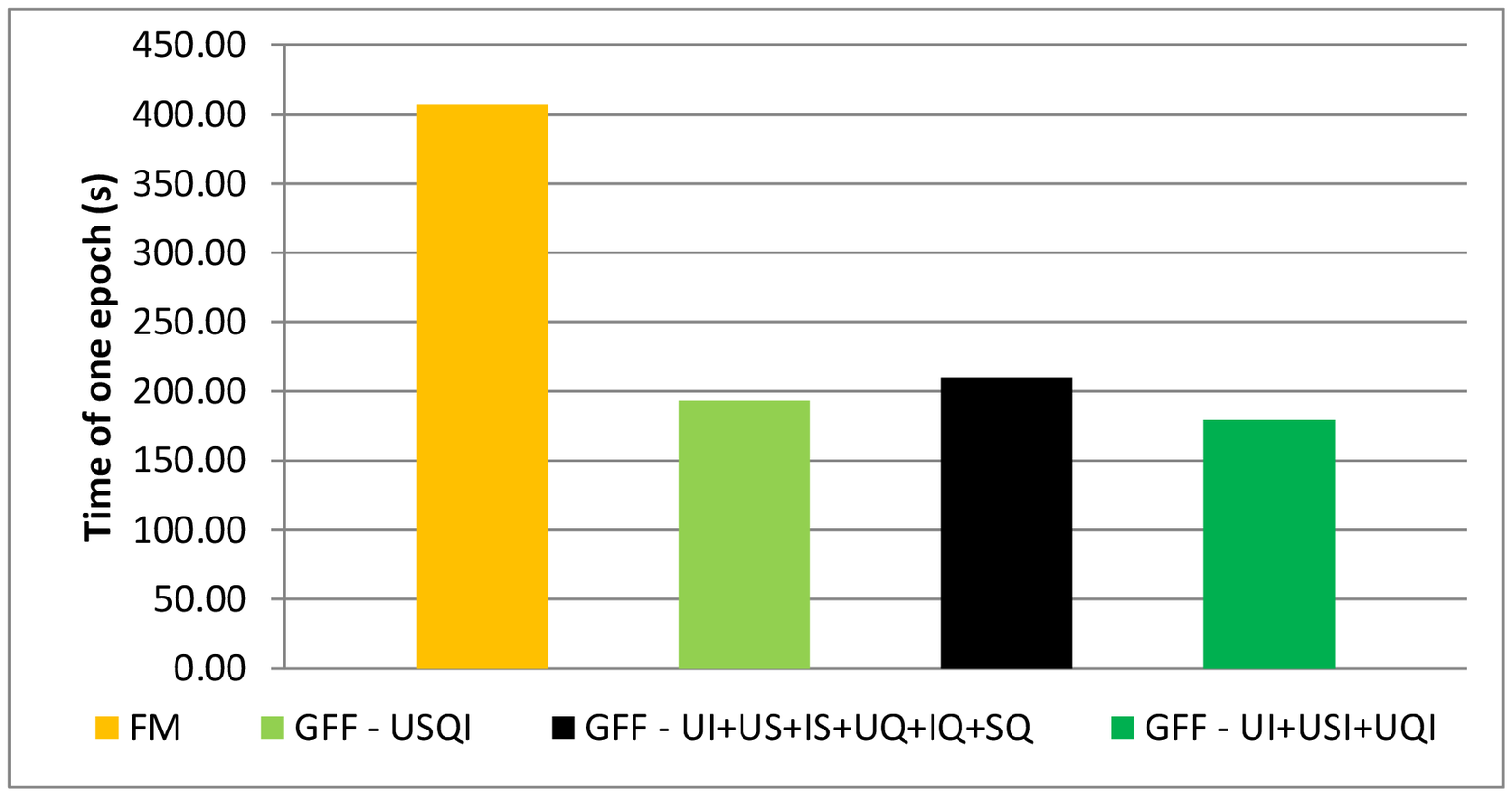}
\caption{Training times of FM and GFF models on the LastFM dataset.}
\label{fig:comptimes}
\end{figure*}

W.r.t. running times we compared FM and GFF. BPR was not included because it does not deal with context and therefore has an unfair advantage. The training time of FM was measured by both libFM's inner logging as well as from external code and the two values were very similar. For this measurement we did not provide a test set for libFM in order to exclude the computation of the test error. Figure~\ref{fig:comptimes} depicts the results on the LastFM dataset. GFF was twice as fast with the pairwise model and even faster with the interaction model. Due to the need of subsampling the negative feedback, FM trains on twice as much examples for the same problem. This increases the time required for training significantly. Note that the results were achieved on a single core of a multi-core CPU, and the training times of GFF can be greatly reduced if multiple cores are used in parallel.

\section{Extension -- MDM compliant GFF}\label{sec:gff-extended}
In this section we lift the restrictions imposed by SA-MDM and extend GFF to allow more attributes per dimensions and thus make it fully compliant with the Multidimensional Dataspace Model. More attributes per dimensions are useful for including multi-value properties of the interacting entities, e.g. tags associated with the items; session behavior; the social network of the users; etc. Such information could be also included in SA-MDM through several dimensions with a single binary attribute. Each attribute describes if a property (e.g.\ a tag) applies to the entity (e.g.\ item) that participates in the transaction. The main drawback of this method that it results in many dimensions and therefore significantly increases training times.

In our solution we intend to handle properties of entities together. This is achieved by bundling their binary attributes into one dimension in accordance with MDM. This admittedly restricts the space of possible preference models by excluding interactions between these attributes. Analogously as for context interactions, we can also argue to exclude property interactions -- between properties of the same kind, since they are irrelevant from the recommendation point of view.

Our solution is inspired by NSVD1 \citep{paterek2007improving} and is as follows.
\begin{enumerate}
\item A dimension should be defined with entities (i.e. different values of the context variable) that are associated with the properties;
\item Each property is represented by an attribute whose value denotes the strength of the attribute for a given entity. A (sparse) mixing matrix ($W\in\mathbb{R}^{S^{(P)}\times S^{(E)}}$, where $S^{(P)}$ and $S^{(E)}$ is the number of properties and entities, respectively) is formed from the values of the attributes.
\item A feature vector is assigned to each property.
\item Since each entity is the weighted sum of its properties, the feature vector of an entity is a weighted sum of the feature vectors of its properties' feature vectors. This allows the learning algorithm to be unchanged for dimensions with single attributes, because the feature vectors can be computed for the entities that directly participate in the transaction. The feature vectors of the entities can be computed using matrix multiplication: $M^{(E)}=M^{(P)}W$.
\end{enumerate}

Since the derivative of the loss function w.r.t.\ the properties' features is not linear in the columns of $M^{(P)}$, an approximative solution is required. We chose to update the properties' feature vectors as if they were independent. To ensure convergence, after training some of the properties' feature vectors, the model should be updated before continuing. Since the update is fast, it can be done after the computation of each vector. Moreover, the update of feature vectors can be parallelized. This method can be still slow if the average number of properties assigned to entities is high.

An other way is to apply two-phase learning similarly to \citep{PilaRecsys09}. The first phase computes $M^{(E)}$ using a normal ALS step. In the second phase $M^{(P)}$ is computed from $M^{(E)}$ and $W$. The finishing step is to compute $M^{(E)}=M^{(P)}W$ from the new $M^{(P)}$, thus the following ALS steps remain consistent. Naturally, the two-phase learning is less accurate, therefore we stick to the direct optimization when possible.

Two examples are shown below on how this extension can be used.

\subsection{Item metadata as attributes}
CBF is often combined with CF to create hybrid algorithms that outperform both of them. E.g.\ item metadata helps overcoming the item cold-start problem in CF \citep{burke2007hybrid}. Here we show how to include item metadata into a model using the extended GFF.

Let us assume that the relevant item metadata is tokenized, preprocessed. From there the outline of the solution is followed.
\begin{enumerate}
\item We create an \emph{item dimension}, its entities are the items, to which we will assign the metadata attributes.
\item Each metadata token is represented by an attribute that indicates the strength of a token for the items. If the item is not associated with the token, the value of the attribute is set to 0 for that item. $W$ is created from these values.
\item A feature vector is assigned to each token.
\item The feature vectors of the items now can be computed as $M^{(I)}=M^{(M)}W$, where $M^{(M)}$ is the feature matrix of the metadata attributes.
\end{enumerate}

\subsection{Session information}
Different sessions of the same user are usually treated uniformly by recommender systems, assuming that user preference does not change across sessions. Session information, however, can be of great help in identifying what the user is currently interested in. This information can further refine recommendations and is exceptionally useful in domains where users likely have broader interests (e.g.\ e-commerce, news sites).

As the context of the transaction, let us assign all items visited during the session but the actual one. Thus the whole session is assigned to each transaction. We exclude the actual item from the session context, since this is the prediction target. Following the outline:
\begin{enumerate}
\item Each transaction will be a separate entity, thus the dimension will consists of all of the transactions. The sessions can not be used as entities, because the associated attributes are different by each transaction of the session since the actual item is omitted.
\item Each item in assigned with an attribute. The attribute is either binary (i.e. the item belongs to the session or not) or weighted by the occurrences of the item in the session. $W$ is created assigning items to each event.
\item Each item is assigned with a feature vector.
\item The feature vectors of the events now can be computed as $M^{(E)}=M^{(X)}W$, where $M^{(X)}$ is the feature matrix of the items. Note that $M^{(X)}$ is a different matrix than the feature matrix of the item dimension $M^{(I)}$.
\end{enumerate}

\subsection{Experimental evaluation}
Initial experiments were done with the extended GFF to incorporate item metadata and session information into the factorization model. The general settings are the same as in Section~\ref{sec:expsetup}. A user session is defined as a sequence of events of a user where the largest gap between two consecutive timestamp is less than 20 minutes. Item metadata consists of the tokenized title, description and category string of the items. The data was filtered for too common and rare tokens. For both context, the weights were $\ell_2$ normalized on an entity by entity basis. The experiments were run on the Grocery dataset, because here the usage of sessions is justified and we have the necessary metadata available.

Session context is denoted by $X$, metadata is by $M$ in our simplified notation. The following models were compared to the classic CF model ($UI$):
\begin{itemize}[noitemsep]
\item \textbf{$\boldsymbol{XI}$:} Interactions between items and the session. Basically this model guesses the actual item based on the other items in the session.
\item \textbf{$\boldsymbol{UI+XI}$:} The classic user--item interaction refined by the actual session.
\item \textbf{$\boldsymbol{UM}$:} The items are replaced by the sum of their metadata in the classic CF model.
\item \textbf{$\boldsymbol{UI+UM}$:} Two aspects of the items are used to model interaction with users, their entity and the sum of their metadata.
\item \textbf{$\boldsymbol{XM}$:} Interaction between other items on the session and the metadata of the actual item.
\end{itemize}

\begin{table}[!h]
\centering
{
\caption{Results for the extended framework on Grocery}\label{tab:extended}
\begin{tabular}{lcc}
\toprule
Model & Recall@20 & Improvement \\
\midrule
$UI$ & 0.1013 & N/A \\
$XI$ & 0.2248 & $+$121.97\% \\
$UI+XI$ & 0.2322 & $+$129.36\% \\
$UM$ & 0.0614 & \hphantom{1}$-$39.34\% \\
$UI+UM$ & 0.2166 & $+$113.87\% \\
$XM$ & 0.2154 & $+$112.77\% \\
\bottomrule
\end{tabular}
}
\end{table}
Table~\ref{tab:extended} summarizes the results. Note that data from the test set is needed for predicting with session in the form of other items of the test session. The results suggest that session information is very important for recommending with the Grocery dataset. $XI$ gives strong result in and of itself and is further improved by mixing in the $UI$ interaction as well. While metadata does not perform well in the place of items, they complements the basic $UI$ model well and is also useful for averting the item cold-start problem.

\section{Conclusion \& future work}\label{sec:conclusion}
In this paper we introduced a general factorization framework, GFF. The novelty of this framework over existing algorithms is its flexibility. It works both on explicit and implicit feedback data, and can incorporate any recommendation context, but even more importantly, the preference model is an input of the algorithm. This allows experimentation with novel preference models without implementing algorithms for every new model separately. The framework optimizes for a weighted square loss function and is very flexible in terms of weighting schemes. This allows us to use the framework for either explicit or implicit feedback based problems and even assumptions on the nature of the missing feedback can be included. The learning is done by a well scaling ALS-CG learner. The computations are smartly decomposed, therefore no sampling of the missing feedback data is required.

In Section~\ref{sec:models} we demonstrated the usefulness of GFF on a four dimensional context-aware recommendation problem. The experimentation showed that certain models capture the preferences of users on items under context much better than the traditional N-way or pairwise models. From the investigated models, we identified the so-called \emph{context interaction model} to be generally useful. This model is the composite of the user--item interaction, refined by context specific user--item interactions ($UI+USI+UQI$ in our simplified notation); to our best knowledge despite its intuitiveness this model was never used before for recommendation. We also found that modeling useless interactions -- such as those between context dimensions -- in fact worsens recommendation accuracy as well as increases the time required for training. The novel models in this framework are generally also more accurate than state-of-the-art algorithms.

GFF was further extended in Section~\ref{sec:gff-extended} to be fully compliant with the Multidimensional Dataspace Model and be able to handle multiple attributes in a dimension. This extension allows for the incorporation of additional data, such as metadata or session information into GFF models. Initial experiments showed that these information can significantly improve the accuracy of the recommendations.

GFF opens up several research paths. While we found a model that works generally well in a common example setting and we had success with the same model is some similar scenarios, the optimal preference modeling for novel tasks is still an open question. Also, we completely ignored models where certain dimensions have multiple sets of features. We think that there is great potential in such models if used properly. A loosely connecting but nonetheless important path is the characterization of context dimensions, i.e.\ determining their quality and their usefulness in $UCI$, $UC$ and $IC$ like interactions prior to training. GFF can help this research by allowing easy evaluation of different context dimensions and models.

GFF can be also improved. We would like to generalize a pairwise ranking loss function and allow for its optimization as an alternative of the current pointwise ranking loss while maintaining the efficiency and scaling of the training. Another potential improvement could be a meta-learner over GFF suggesting the best model for a context-aware recommendation problem and refining it during the training. 

\section*{Acknowledgements}
The work leading to these results has received funding from the European Union's Seventh Framework Programme (FP7/2007-2013) under CrowdRec Grant Agreement n$^\circ$ 610594.

\bibliographystyle{spbasic}      
\bibliography{citations}   

\end{document}